\documentclass[english,letterpaper,aps,preprint]{revtex4}
\usepackage[T1]{fontenc}
\usepackage[latin1]{inputenc}
\usepackage{amsmath}
\usepackage{graphicx}
\usepackage{amssymb}

\makeatletter
\newenvironment{lyxlist}[1]
{\begin{list}{}
{\settowidth{\labelwidth}{#1}
 \setlength{\leftmargin}{\labelwidth}
 \addtolength{\leftmargin}{\labelsep}
 }}
{\end{list}}

\makeatletter

\usepackage{geometry}

\makeatother

\usepackage{babel}
\makeatother
\begin{document}

\title{Dufour and Soret effects in a magnetized and non-magnetized plasma}

\author{L. S. García-Colín}

\address{Departamento de Física, Universidad Autónoma Metropolitana-Iztapalapa,
Av. Purísima y Michoacán S/N, México D. F. 09340, México. Also at
El Colegio Nacional, Luis González Obregón 23, Centro Histórico, México
D. F. 06020, México.}

\author{A. L. García-Perciante}

\address{Depto. de Matemáticas Aplicadas y Sistemas, Universidad Autónoma
Metropolitana-Cuajimalpa, Av. Pedro A. de los Santos No. 84, México
D. F, México}

\author{A. Sandoval-Villalbazo}

\address{Departamento de Física y Matemáticas, Universidad Iberoamericana,
Prolongación Paseo de la Reforma 880, México D. F. 01210, México.}

\date{\today{}}

\begin{abstract}
It is well known that magnetic fields affect heat conduction in a
different way in the direction parallel and perpendicular to the field.
In this paper a formal derivation of this phenomenon and analytical
expressions for the transport coefficients based in the Boltzmann
equation are presented. Moreover, the Dufour effect or diffusion thermo
effect is usually ignored in plasma transport theory. In this work,
analytic expressions for the parallel and perpendicular thermal conductivities
as well as the coefficients for both the thermal diffusion, or Soret,
effect and the Dufour effect are formally derived. It is also shown
how the heat conduction in the perpendicular direction decreases with
increasing magnetic field and how in both directions the diffusion
thermo effect is significant compared to thermal conduction. Other
aspects of this work are also emphasized. 
\end{abstract}
\maketitle

\section{introduction}

The study of transport processes in plasmas is about to reach its
seventieth birthday. Indeed, in 1936-1937 Landau published his pioneering
paper by casting the Boltzmann collision integral in the form of a
Fokker-Planck equation thus regarding that collisions in plasmas may
be treated diffusively. Using very intuitive arguments he then arrived
to what is today known as the collision integral in the logarithmic
approximation for a gas with Coulomb interactions \cite{key-1}. This
method has become the basis of practically all approaches to the study
of such processes in many applications. For instance, most of the
work involving astrophysical problems use this method as further developed
by Spitzer and collaborators \cite{key-2,key-3} over fifty years
ago, also later by other authors \cite{key-4,key-5} and independently
by Braginski \cite{key-7,key-6}. The last effort along these lines
is the apparently not fully recognized but magnificent work on transport
theory in plasmas written by Balescu in 1988 \cite{key-29}. In our
opinion it is probably the best and most complete report written on
the subject as of today. After Balescu's work, research on the kinetic
theory of plasmas seems to have diverted in two directions. One is
related to the kinetic problems which arise in fusion-related research.
In tokamaks and other devices, the geometry and inhomogeneities present
in the relatively large magnetic fields that prevail, affect the particle
dynamics in a crucial way. This topic is referred to as neoclassical
transport \cite{key-30} and the literature here is abundant, a sample
of which may be found in Refs. \cite{key-9,key-10} and the works
cited therein. The second application is characteristic in astrophysical
problems where the Spitzer-Braginski approach seems to dominate. Typical
situations are related to the problem of thermal conduction in tangled
magnetic fields \cite{key-39}, thermal conduction in clusters of
galaxies \cite{key-40} and similar problems \cite{key-41}\cite{key-42}.
A few papers deal with traditional kinetic theory but they do not
seem to proliferate. An example of a formal kinetic treatment of a
two temperature gas can be found in Ref. \cite{key-33} and of a multispecies
plasma in Ref. \cite{key-36}. However, a common denominator in most
recent works in plasma kinetic theory is the use of the Landau-Fokker-Planck
version of the kinetic equation and few have used the full Boltzmann
equation to study this question. Indeed, in a recent authoritative
monograph on the subject \cite{key-43} it is clearly stated that
Braginski's equations {}``are now being universally accepted and
because of their completeness and correctness are considered standard''.
We shall come back to these statements later on.

On the other hand, another approach to the study of transport phenomena
in plasmas, using the full properties of the Boltzmann equation, was
first attempted by Chapman and Cowling in 1939 in the first edition
of their monumental work on kinetic theory \cite{key-11}. It is very
likely that since they had in mind mostly applications to the physics
of metals, their method did not permeate to other branches of physics
as smoothly as the Spitzer-Braginski ones did. In 1960 W. Marshall
wrote a series of three reports entitled {}``The Kinetic Theory of
an Ionized Gas'' in which he extended the preliminary results of
Chapman and Cowling to the calculation of the transport coefficients
of such a system when no magnetic field is present and in the presence
of a magnetic field. Apparently these papers were never published
in a regular journal and have thus been ignored in most work done
in this subject \cite{key-12}.

Another question that has been completely ignored in the several approaches
to the transport processes, with the sole exception of Balescu's work
\cite{key-29}, is the derivation of the basic assumptions of Classical
Irreversible Thermodynamics (CIT) from a kinetic model. This not just
means discussions about the approach to equilibrium of a charged system,
the H-theorem, and the explicit calculation of the entropy production
exhibiting consistency with the second law of thermodynamics. It also
includes the derivation of the complete equations relating the appropriate
fluxes with their corresponding thermodynamic forces, the calculation
of transport coefficients involved therein, specially those associated
with the so-called cross-effects, and then showing explicitly that
the matrix of such transport coefficients is a symmetric one, the
well known Onsager reciprocity theorem. Curiously enough, even for
an inert multicomponent mixture of dilute gases this latter property
was obtained only recently \cite{key-13}. This complete program has
been accomplished for the so-called {}``vectorial'' transport processes
in a dilute plasma, heat conduction, diffusion and electrical conduction.
However the complete work, which includes viscomagnetism, is simply
too bulky to attempt a publication in a single regular article \cite{key-14}.

In this paper we would like to discuss with certain detail the problems
of heat conduction and diffusion using the full Boltzmann equation
as a formalism which not only involves the full two body dynamics
of the interacting species but it also allows one to fulfill the program
described in the previous paragraph. One reason for doing so is that
these processes have become rather important in many astrophysical
applications, specially in the enigmatic problem of the so-called
cooling flows \cite{key-15,key-17,key-40,key-41}. The main result
which we want to emphasize here is that besides the heat transfer
rates which arise from simple heat conduction due to thermal gradients,
results which are essentially in agreement with those of Spitzer and
Braginski, the influence of diffusion (mass transport) is rather surprising.
With or without the presence of a magnetic field, the diffusion thermo
effect or Dufour effect, as well as the pressure thermal effect, seem
to play a rather important role in enhancing heat conduction. To our
knowledge this particular result has never been published. It is as
striking as the thermal analog of the well known Hall effect, the
so-called Righi-Leduc effect, which we have reported in a separate
paper \cite{key-18}. Although similar results are also present in
the case of electrical conduction we shall not deal with them here
to keep the size of this paper within a reasonable length.

Yet, another reason for carrying out this work is to provide the kinetic
foundations of magnetohydrodynamics within the framework of CIT which,
as mentioned above, was partially done in Ref. \cite{key-29} and
later on extended to the relativistic regime \cite{key-44}.

The structure of the paper is as follows. In Section II we briefly
summarize the tenets of classical irreversible thermodynamics, Section
III is devoted to the kinetic theory behind the calculation. In Section
IV we compute and discuss the transport coefficients and Section V
is left for some concluding remarks.

\section{thermodynamic background\label{sec:thermodynamic-background}}

Classical irreversible thermodynamics \cite{key-19,key-20} establishes
that for any physical system in a non-equilibrium state the conservation
equations which govern the time evolution of the corresponding local
state variables together with the assumption that the local entropy
is a time independent functional of the subset of such variables,
which are even under time reversal, lead to the well known entropy
balance equation, namely\begin{equation}
\frac{\partial\left(\rho s\right)}{\partial t}+\nabla\cdot\vec{J}_{s}=\sigma\,.\label{1}\end{equation}
 In Eq.\,(\ref{1}) $\rho$ and $s$ are the local density and entropy
respectively, $\vec{J}_{s}$ is a vector determining the entropy flux
through the system's boundaries and $\sigma$ the entropy production
which may be expressed as a sum of products of the fluxes $\underbar{J}_{i}$
produced by the thermodynamic forces $\underbar{X}_{i}$ (both quantities
being, in general, tensors)\begin{equation}
\sigma=\sum\underbar{J}_{i}\circ\underbar{X}_{i}\,,\label{2}\end{equation}
 where {}``$\circ$'' indicates a product of equal rank tensors.

To supply the sufficient information so that the conservation equations
are well determined one assumes that linear constitutive equations
hold true\begin{equation}
\underbar{J}_{k}=\sum_{j}L_{kj}\underbar{X}_{k}\,,\label{3}\end{equation}
 where the matrix $L_{kj}$, the transport coefficients matrix, must
be symmetrical $L_{kj}=L_{jk}$, the well known Onsager's reciprocity
theorem.

For the particular situation to be considered in this paper, a binary
mixture of an inert ionized gas, Eq.\,(\ref{3}) would be normally
written as follows \cite{key-21}. If $\vec{J}_{q}$ and $\vec{J}_{m_{i}}$
denote the heat and mass fluxes then\begin{equation}
\vec{J}_{q}=-L_{11}\nabla T-L_{12}\vec{d}_{ij}\,,\label{4}\end{equation}
 \begin{equation}
\vec{J}_{i}=-L_{21}\nabla T-L_{22}\vec{d}_{ij}\,,\label{5}\end{equation}
 where $\vec{d}_{ij}=-\vec{d}_{ji}$ is the well known diffusive force.
The coefficient $L_{11}$ is the thermal conductivity, $L_{12}$ is
related to the diffusion thermo effect, $L_{22}$ is the generalized
diffusion coefficient and $L_{21}$ the thermal diffusion coefficient
well known as the Soret effect. More details about them will be discussed
later on. In the $\vec{d}_{ij}$ representation it has recently been
shown that for an ordinary multicomponent ideal mixture \cite{key-13}\begin{equation}
L_{12}=L_{21}\,,\label{6}\end{equation}
 which is Onsager's theorem. This theorem, as pointed out in Ref.\,\cite{key-13}
does not hold for arbitrary choices of the thermodynamic forces but
we shall not pursue this matter here.

The main objective of this paper is to derive Eqs.\,(\ref{4}) and
(\ref{5}) for the case of the two component mixture present in an
ionized gas, namely electrons and ions, in the presence of a magnetic
field with explicit expressions for the transport coefficients and
to carefully analyze their significance. The proof of Onsager's reciprocity
theorem follows at once from the results of Ref. \cite{key-13} in
the absence of a magnetic field. With a magnetic field the proof will
be given elsewhere.

\section{kinetic theory}

In order to carry out the above program we make use of the Boltzmann
equation, implying that our system is a dilute mixture of electrically
charged particles with charges $e_{i}$, masses $m_{i}$ ($i=a,\, b$)
and assume that the ions are protons ($Z=1$). If $f_{i}$ is the
single particle distribution function of species $i$ defined as usual
in $\mu-$space ($\vec{r},\,\vec{v}_{i}$) then,\begin{equation}
\frac{\partial f_{i}}{\partial t}+\vec{v_{i}}\cdot\frac{\partial f_{i}}{\partial\vec{r}}+\frac{e_{i}}{m_{i}}\left(\vec{E}+\vec{v_{i}}\times\vec{B}\right)\cdot\frac{\partial f_{i}}{\partial\vec{v_{i}}}=\sum_{i,j=a}^{b}J\left(f_{i}f_{j}\right)\,.\label{7}\end{equation}
 Here $J\left(f_{i}f_{j}\right)$ is the well known collision kernel
\cite{key-11,key-12} taking into account collisions between particles
of different and same species. Also, to avoid effects arising from
the long range Coulomb potentials, the integration limits are calculated
with a cutoff at a distance $\lambda_{d}$, the Debye cut-off distance
which is the well known value of the impact parameter allowed for
the definition of a collision.

The electromagnetic force is introduced as the Lorentz force where
the magnetic field is weak enough such that the prevailing Larmor
frequency, namely $\omega_{i}=e_{i}B/m_{i}$, is such that $\omega_{i}\tau$
(where $\tau$ is the characteristic time between collisions) is not
to large. Thus, we are considering a regime in which collisions prevail
over cyclotron motion. In this force, the electric and magnetic fields
are the self-consistent fields derived from the Maxwell equations
\cite{key-29}.

Once Eq.\,(\ref{7}) is clearly defined one ought to proceed with
the derivation of the conservation equations and the proof of the
H-theorem. The former is a standard step widely discussed in the literature
\cite{key-11,key-22,key-23}. The second one is a little bit more
tricky due to the presence of the magnetic field \cite{key-14} but
leads, for a wide class of external potentials, to an equilibrium
state described by a Maxwellian distribution. Here, for simplicity,
we are assuming that $T_{a}=T_{b}\equiv T$. Also it is quite clear
that the solution to the homogeneous part of Eq.\,(\ref{7}) is given
by\begin{equation}
f_{i}^{(0)}\left(\vec{r},\vec{v_{i}},t\right)=n\left(\vec{r},t\right)\left(\frac{m_{i}}{2\pi kT\left(\vec{r},t\right)}\right)^{3/2}e^{-\frac{m_{i}c_{i}^{2}}{2kT\left(\vec{r},t\right)}}\,,\label{max}\end{equation}
 where $\vec{c}_{i}=\vec{v}_{i}-\vec{u}\left(\vec{r},t\right)$, $\vec{u}\left(\vec{r},t\right)=\left(n_{a}\vec{u}_{a}+n_{b}\vec{u}_{b}\right)/n$
are the chaotic and barycentric velocity respectively, $n=n_{a}+n_{b}$
is the number density and all these quantities are locally defined.
The local temperature $T\left(\vec{r},t\right)$ is defined through
the internal energy density $\varepsilon\left(\vec{r},t\right)$ by
the standard ideal gas law. In Eq.\,(\ref{max}) the electric field
has been ignored for simplicity, as will be in the rest of this article
but can be easily incorporated in the formalism through its potential
as a conservative force.

We thus proceed directly to the solution of Eq.\,(\ref{7}) following
the Hilbert-Chapman-Enskog method widely discussed in the literature
\cite{key-11,key-22,key-23}. The method is based on two assumptions,
firstly that $f_{i}$ can be expanded in a power series in Knudsen's
parameter $\epsilon$ which is a measure of the magnitude of the macroscopic
gradients within the extension of the physical system and secondly,
that the time dependence of $f_{i}$ occurs only through the local
state variables, $n\left(\vec{r},t\right)$, $\vec{u}\left(\vec{r},t\right)$
and $\varepsilon\left(\vec{r},t\right)$, or equivalently $T\left(\vec{r},t\right)$.
Then\begin{equation}
f_{i}\left(\vec{r},\vec{v_{i}},t\right)=f_{i}^{(0)}\left(\vec{r},\vec{v_{i}}|\, n,\vec{u},T\right)\left[1+\epsilon\varphi_{i}^{\left(1\right)}\left(\vec{r},\vec{v_{i}}|\, n,\vec{u},T\right)+\mathcal{O}\left(\epsilon^{2}\right)\right]\,,\label{9}\end{equation}
 where $n,$ $\vec{u}$ and $T$ are functions of $\vec{r}$ and $t$.
For $\epsilon=0$, $f_{i}=f_{i}^{(0)}$ which when substituted in
the conservation equations leads to the Euler equations of magnetohydrodynamics.
To first order in $\epsilon$, since the right side of Eq.\,(\ref{7})
is already of order $\epsilon$ we get, after substitution of Eq.\,(\ref{9})
in Eq.\,(\ref{7}) and a lengthy but straightforward manipulation
of the left side, the linearized Boltzmann equation, namely\begin{eqnarray}
\frac{m_{i}}{kT}\vec{c}_{i}{}^{^{^{^{0}}}}\!\!\vec{c}_{i}:\nabla^{^{^{0}}}\!\!\vec{u}+\left[\left(\frac{m_{i}c_{i}^{2}}{2kT}-\frac{5}{2}\right)\frac{\nabla T}{T}+\frac{n}{n_{i}}\vec{d}_{ij}\right]\cdot\vec{c}_{i} & = & -\frac{m_{i}}{\rho kT}\sum_{j=a}^{b}e_{j}\int d\vec{c}_{j}\, f_{j}^{(0)}\varphi_{j}^{(1)}\left(\vec{c}_{j}\times\vec{B}\right)\cdot\vec{c}_{i}\nonumber \\
 &  & \hspace{-15em}-\frac{e_{i}}{m_{i}}\left(\vec{c}_{i}\times\vec{B}\right)\cdot\frac{\partial\varphi_{i}^{(1)}}{\partial\vec{v}_{i}}+C\left(\varphi_{i}^{(1)}\right)+C\left(\varphi_{i}^{(1)}\varphi_{j}^{(1)}\right)\quad i=a,\, b\,,\label{10}\end{eqnarray}
 where the superscript {}``$0$'' over a tensor indicates its symmetric
and traceless part. In Eq.\,(\ref{10}), $C\left(\varphi_{i}^{(1)}\right),\, C\left(\varphi_{i}^{(1)}\varphi_{j}^{(1)}\right)$
are the linearized collision kernels whose explicit form is also well
known \cite{key-11,key-22,key-23} and $\vec{d}_{ij}$ is the diffusive
vector force given by\begin{equation}
\vec{d}_{ij}\equiv\nabla\left(\frac{n_{i}}{n}\right)+\frac{n_{i}n_{j}}{n}\left(\frac{m_{j}-m_{i}}{\rho}\right)\frac{\nabla p}{p}+\frac{n_{i}n_{j}}{\rho p}\left(m_{i}e_{j}-m_{j}e_{i}\right)\vec{u}\times\vec{B}\,,\label{11}\end{equation}
 which satisfies the property $\vec{d}_{ij}=-\vec{d}_{ji}$.

The solution to Eq.\,(\ref{10}) follows from the standard technique
in the theory of integral equations using the fact that its structure
resembles that of a system of linear inhomogeneous integral equations.
Neglecting the term $\frac{m_{i}}{kT}\vec{c}_{i}{}^{^{^{^{0}}}}\!\!\vec{c}_{i}:\nabla^{^{^{0}}}\!\!\vec{u}$,
which involves second rank tensors and is important only when studying
viscomagnetic effects \cite{key-14}, one gets that (see Refs.\,\cite{key-13}
and \cite{key-22}),\begin{equation}
\varphi_{i}^{(1)}=\vec{\mathbb{A}}_{i}\cdot\nabla T+\vec{\mathbb{D}}_{i}\cdot\vec{d_{ij}}\,,\label{12}\end{equation}
 where $\vec{\mathbb{A}}_{j}$ and $\vec{\mathbb{D}}_{i}$ are the
most general vectors that can be constructed from the base vectors
in our space $\vec{c}_{i}$, $\vec{c_{i}}\times\vec{B}$ and $\left(\vec{c_{i}}\times\vec{B}\right)\times\vec{B}=\vec{B}\left(\vec{c}_{j}\cdot\vec{B}\right)-B^{2}\vec{c}_{j}$,
whence\begin{equation}
\vec{\mathbb{A}}_{i}=\mathbb{A}_{i}^{\left(1\right)}\vec{c}_{i}+\mathbb{A}_{i}^{\left(2\right)}\vec{c}_{i}\times\vec{B}+\mathbb{A}_{i}^{\left(3\right)}\vec{B}\left(\vec{c}_{i}\cdot\vec{B}\right)\,,\label{13}\end{equation}
 and\begin{equation}
\vec{\mathbb{D}}_{i}=\mathbb{D}_{i}^{\left(1\right)}\vec{c}_{i}+\mathbb{D}_{i}^{\left(2\right)}\vec{c}_{i}\times\vec{B}+\mathbb{D}_{i}^{\left(3\right)}\vec{B}\left(\vec{c}_{i}\cdot\vec{B}\right)\,,\label{14}\end{equation}
 where $\mathbb{A}_{i}^{\left(k\right)}$, $\mathbb{D}_{i}^{\left(k\right)}$
($k=1,\,2,\,3$) are functions of all scalars in this space, ($n,\, T,\, c_{i}^{2},\, B^{2},\,\left(\vec{c}_{i}\cdot\vec{B}\right)^{2}$).

The next step is to substitute Eqs.\,(\ref{13}) and (\ref{14})
into Eq.\,(\ref{10}) and separate terms in $\nabla T$ and $\vec{d}_{ij}$.
The algebraic procedure is both lengthy and tricky. It is outlined
in Appendix A. The final result is that one gets a set of two linear
integral equations for each of the functions $\vec{\mathbb{A}}_{i}$
and $\vec{\mathbb{D}}_{i}$ namely\begin{equation}
f_{i}^{(0)}\left(\frac{mc_{i}^{2}}{2kT}-\frac{5}{2}\right)\vec{c_{i}}=f_{i}^{(0)}\left\{ C\left(\vec{c}_{i}\mathcal{R}_{i}\right)+C\left(\vec{c}_{i}\mathcal{R}_{i}+\vec{c}_{j}\mathcal{R}_{j}\right)\right\} \,,\label{15}\end{equation}
 \begin{eqnarray}
f_{i}^{(0)}\left(\frac{mc_{i}^{2}}{2kT}-\frac{5}{2}\right)\vec{c_{i}} & = & -f_{i}^{(0)}\frac{m_{i}}{kT}iB\vec{c}_{i}\mathcal{G}-f_{i}^{(0)}\frac{e_{i}}{m_{i}}\vec{c}_{i}iB\mathcal{A}_{i}\nonumber \\
 & + & f_{i}^{(0)}\left\{ C\left(\vec{c}_{i}\mathcal{A}_{i}\right)+C\left(\vec{c}_{i}\mathcal{A}_{i}+\vec{c}_{j}\mathcal{A}_{j}\right)\right\} \,,\label{16}\end{eqnarray}
 where $\mathcal{R}_{i}=\mathbb{A}_{i}^{\left(1\right)}+B^{2}\mathbb{A}_{i}^{\left(3\right)}$
for $i=a,\, b$. Also, in Eq. (\ref{16})\begin{equation}
\mathcal{A}_{i}=\mathbb{A}_{i}^{\left(1\right)}+iB\mathbb{A}_{i}^{\left(2\right)}\,,\label{17}\end{equation}
 \begin{equation}
\mathcal{G}=G_{B}^{\left(1\right)}+iBG_{B}^{\left(2\right)}\,,\label{18}\end{equation}
 where $G_{B}^{\left(1\right)}$ and $G_{B}^{\left(2\right)}$ are
defined in Appendix A.

An identical set of equations is obtained for $\mathbb{D}_{i}^{(k)}$
except that the inhomogeneous term is $f_{i}^{(0)}\frac{n}{n_{i}}\vec{c_{i}}$,
the $\mathcal{A}_{i}$'s are substituted by $\mathcal{D}_{i}$'s and
$\mathcal{G}$ by a function $\mathcal{K}$ also defined in the Appendix
A. Equations (\ref{15}) and (\ref{16}) and their analogs for the
$\mathcal{D}_{i}$ functions are the furthest results that can be
obtained without resorting to the explicit form of the interaction
potential between the different species. The evaluation of the linearized
collision terms involved in the homogeneous parts of these equations
will lead to the standard collision integrals \cite{key-12,key-21}.

The solution to Eqs.\,(\ref{15}) and (\ref{16}) is somewhat lengthy.
We fist notice that Eq.\,(\ref{15}) does not contain any explicit
contribution of the magnetic field, the $\mathcal{R}_{i}$ function
does not involve any property other than those already associated
with the $\mathbb{A}_{i}$'s so that its solution can be immediately
implied from the one arising for the same problem involving only an
inert dilute mixture and its solution is available in any standard
book on kinetic theory \cite{key-12,key-22,key-23}, although we will
come back to it in Appendix B. On the other hand Eq.\,(\ref{16})
is different on account of the first two terms in the right side involving
the magnetic field ($iB$). To accomplish its solution we follow the
almost orthodox methods in kinetic theory, namely to expand the unknown
functions $\mathbb{A}_{i}^{\left(k\right)}$ (and $\mathbb{D}_{i}^{\left(k\right)}$)
in terms of a complete orthonormal set of functions, the Sonine (Laguerre)
polynomials. Thus we write that\begin{equation}
\mathbb{A}_{j}^{(k)}=\sum_{m=0}^{\infty}a_{j}^{(k)(m)}S_{3/2}^{(m)}\left(c_{i}^{2}\right)\qquad k=1,\,2,\,3\,,\label{19}\end{equation}
 where the Sonine polynomials are defined elsewhere \cite{key-21,key-23}
and have the property that\begin{equation}
\int_{0}^{\infty}e^{-x^{2}}S_{n}^{(p)}(x)\, S_{n}^{(q)}(x)\, x^{2n+1}dx=\frac{\Gamma(n+p+q)}{2p!}\delta_{pq}\,,\label{20}\end{equation}
 and the first few of them are $S_{n}^{(0)}(x)=1$, $S_{n}^{(1)}(x)=-x+n+1$
and\begin{equation}
S_{n}^{(2)}(x)=\frac{1}{2}\left(n+1\right)\left(n+2\right)-\left(n+2\right)x+\frac{x^{2}}{2}\,.\label{21}\end{equation}
 In Eq.\,(\ref{19}) the coefficients $a_{j}^{(k)(m)}$ are functions
of the scalars $n,\, T,\, c_{i}^{2},\, B^{2}$, and $\left(\vec{c}_{i}\cdot\vec{B}\right)^{2}$.
A similar expression holds for $\mathbb{D}_{j}^{(k)}$.

Clearly, for Eq.\,(\ref{17}),\begin{equation}
\mathcal{A}_{i}=\mathbb{A}_{i}^{\left(1\right)}+iB\mathbb{A}_{i}^{\left(2\right)}=\sum_{m=0}^{\infty}a_{j}^{(m)}S_{3/2}^{(m)}\left(c_{i}^{2}\right)\,,\label{22}\end{equation}
 with\begin{equation}
a_{i}^{(m)}=a_{i}^{(1)(m)}+iB\, a_{i}^{(2)(m)}\,.\label{23}\end{equation}

Equations (\ref{19}), (\ref{22}) and (\ref{23}) considerably reduce
both the structure of the integral equation {[}Eq.\,(\ref{16})]
and its analog for the $\mathbb{D}_{i}^{(k)}$ functions as well as
the subsidiary conditions, Eqs.\,(\ref{c1}) in Appendix C. In fact,
introducing the dimensionless velocity\begin{equation}
\vec{w_{i}}=\sqrt{\frac{m_{i}}{2kT}}\vec{c}_{i}\,,\label{24}\end{equation}
 noticing that $f_{i}^{(0)}=n_{i}\left(\frac{m_{i}}{2\pi kT}\right)^{3/2}e^{-w_{i}^{2}}$
and using Eq.\,(\ref{20}) we get, after straightforward integration
that\begin{equation}
\sum_{i=a}^{b}n_{i}\left(a_{i}^{(1)(0)}+B^{2}a_{i}^{(3)(0)}\right)=0\,,\label{25}\end{equation}
 \begin{equation}
\sum_{i=a}^{b}n_{i}a_{i}^{(0)}=0\,,\label{26}\end{equation}
 this implying that all coefficients $a_{i}^{(n)}$ for $n>0$ are
not restricted, and also that\begin{equation}
\mathcal{G}=\sum_{j}\frac{n_{j}e_{j}}{m_{j}}kT\, a_{j}^{(0)}\,,\label{27}\end{equation}
 and a similar expression for the function $\mathcal{K}$ defined
in the Appendix A.

As we shall see below, the solution to the integral equations (\ref{15})
and (\ref{16}) is considerably simplified if we first compute explicitly
the fluxes in terms of the forces, Eqs.\,(\ref{4}) and (\ref{5}),
and obtain formal expressions for the transport coefficients. This
will be the subject of the following section.

\section{transport coefficients}

This section is devoted to the explicit calculation of both the mass
and heat fluxes using the results of the previous sections and some
rather well known results of irreversible thermodynamics. Let us begin
with the transport of mass. The diffusion flux for species $a$ is
defined as\begin{equation}
\vec{J}_{a}=m_{a}\int\vec{c}_{a}f_{a}^{(0)}\varphi_{a}^{(1)}d\vec{c}_{a}\,.\label{28}\end{equation}
 Substituting Eqs.\,(\ref{11}) and (\ref{12}) into Eq.\,(\ref{28})
and noticing that all the resulting integrals contain the dyad $\vec{c}_{a}\vec{c}_{a}$,
we get with the use of Eqs.\,(\ref{19}) and (\ref{20}) that\begin{equation}
\int f_{a}^{(0)}\sum_{m=0}^{\infty}a_{a}^{(i)(m)}S_{3/2}^{(m)}\vec{c}_{a}\vec{c}_{a}d\vec{c}_{a}=\frac{n_{a}}{m_{a}}kTa_{a}^{(i)(0)}\mathbb{I}\,,\label{29}\end{equation}
 where $\mathbb{I}$ is the unit tensor and use has been made of the
fact that the contribution arising from the symmetric traceless part
of $\vec{c}_{a}\vec{c}_{a}$ vanishes. The same result is obtained
for the $\mathbb{D}_{i}$ functions so that after collecting all terms
we get that,\begin{eqnarray}
\vec{J}_{a} & = & n_{a}k\left\{ a_{a}^{(1)(0)}\nabla T+a_{a}^{(2)(0)}\nabla T\times\vec{B}+a_{a}^{(3)(0)}\left(\vec{B}\cdot\nabla T\right)\vec{B}\right.\nonumber \\
 &  & \left.+T\left[d_{a}^{(1)(0)}\vec{d}_{ab}+d_{a}^{(2)(0)}\vec{d}_{ab}\times\vec{B}+\left(\vec{B}\cdot\vec{d}_{ab}\right)\vec{B}\, d_{a}^{(3)(0)}\right]\right\} \,,\label{30}\end{eqnarray}
 and a similar expression for $\vec{J}_{b}$ since $\vec{J}_{a}+\vec{J}_{b}=0$
and $\vec{d}_{ab}=-\vec{d}_{ba}$. Equation (\ref{30}) is a very
eloquent result namely, knowledge of mass flux requires not the full
information contained in the infinite series of Eq.\,(\ref{19})
but only of six coefficients, three for each of the thermodynamic
forces, $\nabla T$ and $\vec{d}_{ij}$. This fact will considerably
simplify the solution of the integral equations derived in the previous
sections. Notice also how the magnetic field influences the diffusive,
thermal and mechanical effects which give rise to the mass flux. This
is readily appreciated examining Eq.\,(\ref{11}) which even in the
absence of an external conservative field has two contributions, one
arising from the standard diffusion (the first two terms) and another
one arising from the magnetic force term $\vec{u}\times\vec{B}$.
Moreover, the thermal diffusion effect (proportional to $\nabla T$)
is given by the first three terms in Eq.\,(\ref{30}), each one representing
a different contribution. Referring our vectors to a cartesian coordinate
system, if we choose $\vec{B}$ along the $z$-axis, $\vec{B}=B\hat{k}$,
then the third term is simply $B^{2}\frac{\partial T}{\partial z}\hat{k}$
which applies as well to the last term, so that Eq.\,(\ref{30})
may be rewritten as\begin{eqnarray}
\vec{J}_{a} & = & n_{a}k\left\{ \left(a_{a}^{(1)(0)}+B^{2}a_{a}^{(3)(0)}\right)\nabla_{\parallel}T+a_{a}^{(1)(0)}\nabla_{\perp}T+a_{a}^{(2)(0)}B\nabla_{s}T\right.\nonumber \\
 &  & +\left.T\left[\left(d_{a}^{(1)(0)}+B^{2}d_{a}^{(3)(0)}\right)\vec{d}_{ab\parallel}+d_{a}^{(1)(0)}\vec{d}_{ab\perp}T+d_{a}^{(2)(0)}B\vec{d}_{ab\, s}\right]\right\} \,,\label{31}\end{eqnarray}
 where $\nabla_{\parallel}T=\frac{\partial T}{\partial z}\hat{k}$
and such term represents the mass flux parallel to the magnetic field;
$\nabla_{\perp}T=\frac{\partial T}{\partial x}\hat{i}+\frac{\partial T}{\partial y}\hat{j}$
is the thermal force giving rise to the flow of mass in the plane
perpendicular to the magnetic field and finally $\nabla_{s}T=\frac{\partial T}{\partial x}\hat{j}-\frac{\partial T}{\partial y}\hat{i}$
is the thermal force producing a mass flow which is perpendicular
to both the direction of the field and to $\nabla T$. Similar features
also hold for the diffusion force $\vec{d}_{ij}$ which will be analyzed
in detail later on. We must also mention that when $\vec{B}=0$, this
last contribution vanishes and Eq.\,(\ref{31}) reduces to\begin{equation}
\vec{J}_{a}=n_{a}k\, a_{a}^{(1)(0)}\nabla T+n_{a}kT\, d_{a}^{(1)(0)}\vec{d}_{ab}\,,\label{32}\end{equation}
 which is of the standard form required by non-equilibrium thermodynamics
{[}cf. Eq.\,(\ref{5})]. Here $n_{a}k\, a_{a}^{(1)(0)}$ is the so-called
Soret or thermal diffusion coefficient \cite{key-19,key-20}. In the
presence of a magnetic field there will be three contributions to
the Soret effect which are easily written down from Eq.\,(\ref{31})
and the results of Appendix B. Indeed, if we denote these coeficients
by $S$, then we get that\[
S_{\parallel}=\frac{nk\tau}{2}\times2.9\quad\frac{kg}{s\, m\, K}\,,\]
where $\varphi$ is defined in Eq.\,(\ref{b15}). Similarly, using
Eqs.\,(\ref{b3}) and (\ref{aes}) one gets that\[
S_{\perp}=\frac{nk\tau}{2\Delta_{1}}\left(56.3-662x^{2}-2.25x^{4}\right)\quad\frac{kg}{s\, m\, K}\,,\]
where\begin{equation}
\Delta_{1}(x)=19+2078x^{2}+2650x^{4}+9x^{6}\,.\label{48}\end{equation}
 Here $x=\omega_{e}\tau$ where the mean free time $\tau$, is obtained
directly from the collision integrals (see Appendix B), and is given
by\begin{equation}
\tau=\frac{4\left(2\pi\right)^{3/2}\sqrt{m_{e}}\left(kT\right)^{3/2}\epsilon_{0}^{2}}{ne^{4}\psi}\,,\label{49}\end{equation}
 Finally, using the same equations\[
S_{s}=\frac{nk\tau}{{2\Delta}_{1}}\left(647x+2.2x^{3}\right)\quad\frac{kg}{s\, m\, K}\,,\]
 The values of these coefficients as functions of $x$ are plotted
on Fig. 1.

Let us now turn our attention to the flow of heat. According to the
tenets of CIT the heat current is defined as \cite{key-19,key-20}\begin{equation}
\vec{J_{q}'}=\vec{J}_{q}-\sum_{k}h_{k}\frac{\vec{J}_{k}}{m_{k}}\,,\label{33}\end{equation}
 where $h_{k}$ is the enthalpy of the $k^{th}$ component which is
equal to $\frac{5}{2}kT$ for an ideal gas. Thus in kinetic theory,
the analog of this equation is clearly given by\begin{equation}
\vec{J_{q}'}=\vec{J}_{q}-\frac{5}{2}kT\left(\frac{\vec{J}_{a}}{m_{a}}+\frac{\vec{J}_{b}}{m_{b}}\right)\,,\label{34}\end{equation}
 or using the standard definition of $\vec{J}_{q}$ as well as Eq.\,(\ref{28})
we get that,\begin{equation}
\vec{J_{q}'}=kT\sum_{i=a}^{b}\int\left(\frac{m_{i}c_{i}^{2}}{2kT}-\frac{5}{2}\right)\vec{c}_{i}f_{i}^{(0)}\varphi_{i}^{(1)}d\vec{c}_{i}\label{35}\end{equation}
 We now perform the following steps: substitute Eqs.\,(\ref{12}-\ref{14})
and (\ref{19}) in Eq. (\ref{35}) and after examination of the six
resulting terms realize that all of them have as common factor the
integral\begin{equation}
\sum_{i=a}^{b}\int\left(\frac{m_{i}c_{i}^{2}}{2kT}-\frac{5}{2}\right)\frac{1}{3}c_{i}^{2}f_{i}^{(0)}\sum_{m=0}^{\infty}a_{i}^{(j)(m)}S_{3/2}^{(m)}d\vec{c}_{i}=-\frac{5}{2}kT\sum_{i}\frac{n_{i}}{m_{i}}a_{i}^{(j)(1)}\,,\label{36}\end{equation}
 and a similar result for the $d_{i}^{(k)}$ functions. This leads
to the result that\begin{eqnarray}
\vec{J}_{q}' & = & -\frac{5}{2}k^{2}T\sum_{i}\frac{n_{i}}{m_{i}}\left\{ a_{i}^{(1)(1)}\nabla T+a_{i}^{(2)(1)}\vec{B}\times\nabla T+a_{i}^{(3)(1)}\left(\vec{B}\cdot\nabla T\right)\vec{B}\right.\nonumber \\
 &  & +\left.T\left[d_{i}^{(1)(1)}\vec{d}_{ab}+d_{i}^{(2)(1)}\vec{B}\times\vec{d}_{ab}+d_{i}^{(3)(1)}\left(\vec{B}\cdot\vec{d}_{ab}\right)\vec{B}\right]\right\} \,,\label{37}\end{eqnarray}
 where care has to be taken in the second summation to account for
the property that $\vec{d}_{ab}=-\vec{d}_{ba}$. Equation (\ref{37})
contains all possible contributions to the flow of heat including
those arising from the magnetic field since $\vec{u}\times\vec{B}$
is present in the diffusive force $\vec{d}_{ab}$. Moreover, each
term has the same structure of the two contributions appearing in
Eqs.\,(\ref{30}) and (\ref{31}) so that there will be three directions
of heat flow, the one parallel to $\vec{B}$ and which is unaffected
by the field, one perpendicular to $\vec{B}$ and a third perpendicular
to both $\vec{B}$ and $\nabla T$. These latter one is the analog
of the Hall effect in electromagnetism and is known as the Righi-Leduc
effect. This effect, discovered in 1887, was carefully studied by
W. Voigt in 1903 \cite{key-20,key-24} and discussed elsewhere \cite{key-18}.

Clearly, if $\vec{B}=0$ the ordinary (Fourier) thermal conductivity
is simply given by\begin{equation}
\kappa_{\parallel}=\frac{5}{2}k^{2}T\sum_{i=a}^{b}\frac{n_{i}}{m_{i}}a_{i}^{(1)(1)}\,,\label{38}\end{equation}
 whose calculation demands only knowledge of two coefficients $a_{i}^{(1)(1)}$,
$i=a,\, b$, not restricted by subsidiary conditions, instead of the
full knowledge of the infinite series involved in Eq.\,(\ref{19}).
In the diffusion thermo effect, usually referred to as the Dufour
effect, a little care is required to examine the transport coefficient
since even if $\vec{B}=0$, $\vec{d}_{ab}$ contains two contributions,
one arising from $\nabla\left(\frac{n_{a}}{n}\right)$ and another
one, the so-called pressure diffusion coefficient proportional to
$\nabla p$ {[}c.f. Eq.\,(\ref{11})]. The resulting equation is
of the form\begin{equation}
\vec{J}_{q}'=\kappa\nabla T-D_{12}\vec{d}_{ab}\,,\label{39}\end{equation}
 where\begin{equation}
D_{12}=\frac{5}{2}\left(kT\right)^{2}\sum_{i=a}^{b}\frac{n_{i}}{m_{i}}d_{i}^{(1)(1)}\,,\label{40}\end{equation}
 which is of the standard form required by non-equilibrium thermodynamics.

The final step to be undertaken here is to calculate the twelve coefficients
which appear in Eqs.\,(\ref{31}) and (\ref{37}), respectively and
that requires the explicit solution of the integral equations (\ref{15})
and (\ref{16}) and their analogs for the $\mathbb{D}_{i}^{(k)}$
functions under the assumption stated in Eq.\,(\ref{19}) together
with the subsidiary conditions, Eqs.\,(\ref{25}) and (\ref{26})
and their analogs for the $d_{i}^{(0)}$ coefficients. This involves,
once more, a rather lengthy and strictly mathematical procedure including
the detailed evaluation of fourteen collision integrals which for
completeness reasons, is outlined in Appendix B. In what follows we
shall simply quote the results there obtained. Let us concentrate
on the first line of Eq.\,(\ref{37}), set $\vec{B}=B\hat{k}$ along
the $z$-axis and write for the conductive part of the heat flux,\begin{equation}
\left(\vec{J}_{q}'\right)^{c}=-\kappa_{\parallel}\nabla_{\parallel}T-\kappa_{\perp}\nabla_{\perp}T-\kappa_{s}\nabla_{s}T\,,\label{41}\end{equation}
 where each term can be easily identified form Eq.\,(\ref{37}).
Thus, using Eqs.\,(\ref{b15}) and (\ref{alfas}) we obtain,\begin{equation}
\kappa_{\parallel}=\frac{5}{4}\frac{k^{2}T}{m_{e}}\times2.01n\tau\quad\frac{J}{s\, K\, m}\,.\label{42}\end{equation}
As expected, $\kappa_{\parallel}$ does not depend on $\vec{B}$ and
has been obtained taking into account that for $Z=1$, a hydrogen
plasma, $m_{p}=1836m_{e}$ and assuming further that $n_{a}=n_{b}=n/2$,
the plasma is fully ionized. $k$ is Boltzmann's constant, $m_{e}$
and $e$ are the electron mass and charge, respectively, $\epsilon_{0}$
the electrical constant in $\text{Farads}-m^{-1}$ and $\psi$ is
Coulomb's logarithmic function\begin{equation}
\psi=\ln\left[1+\left(\frac{16\pi kT\lambda_{d}\epsilon_{0}}{e^{2}}\right)^{2}\right]\,,\label{43}\end{equation}
 obtained from the collision integrals. Debye's characteristic shielding
length, $\lambda_{d}$, is defined as\begin{equation}
\lambda_{d}=\sqrt{\frac{\epsilon_{0}kT}{ne^{2}}}\,.\label{44}\end{equation}
 It is important to mention that Eq.\,(\ref{42}) is in agreement
with the Spitzer-Braginski result to within 10\% as the interested
reader may easily verify. On the other hand the expressions for $\kappa_{\perp}$
and $\kappa_{s}$ are strongly field dependent and according to the
set of equations (\ref{aes}) are given by\begin{equation}
\kappa_{\perp}=\frac{5}{4}\frac{k^{2}T}{m_{e}}\frac{n\tau}{\Delta_{1}}\times\left(38.7+2270x^{2}+161x^{4}\right)\quad\frac{J}{s\, K\, m}\,,\label{45}\end{equation}
\begin{equation}
\kappa_{s}=\frac{5}{4}\frac{k^{2}T}{m_{e}}\frac{n\tau}{\Delta_{1}}\times\left(206x+2644x^{3}\right)\quad\frac{J}{s\, K\, m}\,,\label{46}\end{equation}
and $\Delta_{1}$ is defined in (\ref{48}). Figure 2 clearly exhibits
the behavior of the three conductivities for weak fields. A number
of comments about these results is pertinent. In the first place $\kappa_{\perp}$,
the thermal conductivity for the heat flow along the $\left(x,\, y\right)$
plane, reduces to $\kappa_{\parallel}$ when $\vec{B}=0$ as may be
easily verified. Since in this case $\kappa_{s}=0$ we recover Eq.\,(\ref{39}),
that is the ordinary Fourier's constitutive equation. These results
for the parallel and perpendicular components of the heat flux (for
$\omega\tau\ll1$) agree with those in Ref. \cite{key-29}. Secondly,
the Righi-Leduc conductivity {[}Eq.\,(\ref{46})] vanishes when $\vec{B}=0$
and for a whole range of values $\vec{B}\neq0$ never exceeds the
value of $\kappa_{\parallel}$. This is in agreement with Balescu's
results \cite{key-29}.

On the other hand, it is also interesting to compare with the original
results obtained by Marshall \cite{key-12}. To do so we must clarify
exactly and precisely where the differences are. We begin with the
equation for the electrical current written at the bottom of page
22 in Marshall's report number 3, which leads to Eq. (3.61). This
correct result is identical to ours. Next, he defines a new diffusive
force $\vec{D}$, \begin{equation}
\vec{D}=\frac{-p\rho}{e_{1}m_{2}-e_{2}m_{1}}\vec{d_{1}}\label{49.5}\end{equation}
 where $\vec{d_{1}}$ is given in Eq. (3.14). So far, so good. Then
he writes the electrical current as shown in Eq. (3.62) and concludes
that the coefficients $\sigma_{I}$, $\sigma_{II}$ and $\sigma_{III}$
are the thermal diffusion coefficients. This is questionable. Firstly,
the force $\vec{D}$ has three components, a coefficient times $\nabla(\frac{n_{1}}{n})$,
a coefficient times $\nabla p$ and the coefficient of $\vec{E'}$,
see Eq. (3.63). The coefficient of $\nabla(\frac{n_{1}}{n})$, a concentration
gradient that produces an electrical current is called the \char`\"{}Dorn
effect\char`\"{}. The coefficients of $\vec{E'}$ are the electrical
conductivities, when we have the case $\nabla p=0$ (the pressure
contribution to $\vec{j}$). The coefficients $\phi^{I}$, $\phi^{II}$
and $\phi^{III}$, a thermal gradient contributing to $\vec{j}$,
are the Thomson coefficients (thermoelectricity) \emph{none} of which
have to do with thermal diffusion. As we shall point out later, mass
and electrical currents are easily related to each other {[}see Eq.
(\ref{59})] whence, except for a numerical factor, $\frac{m_{1}m_{2}e}{m1+m2}$,
the coefficients of $\nabla T$, $\phi^{I}$-$\phi^{III}$, are indeed
the same as the three Soret coefficients which we have explicitly
obtained in a different context (see Eq. (\ref{31}) and Fig. 1 ).
On the other hand, except for a numerical scaling factor, $\sigma^{I}$,
$\sigma^{II}$ and $\sigma^{III}$ (Eqs. 7.8-7.10) are proportional
to the ordinary diffusion coefficients \emph{not} discussed here.

Next we consider the equation for the flow of heat. In the form written
by Marshall in Eq. (7.16), the $\theta^{I}$, $\theta^{II}$ and $\theta^{III}$
coefficients, the thermal conductivities given in Eq. (7.19) are not
precisely our $\kappa_{\parallel}$, $\kappa_{\perp}$ and $\kappa_{s}$
because our definitions of heat flow differ by the term $\frac{5}{2}kT(n_{1}\left\langle \vec{c}_{1}\right\rangle +n_{2}\left\langle \vec{c}_{2}\right\rangle )$
which according to CIT should be included in $\vec{J}'_{q}$ {[}see
Eqs. (\ref{33}) and (\ref{34}) ] whereas Marshall deals with it
\emph{a la} Chapman and Cowling leading to complicated expressions
for the thermal conductivities. Nevertheless the $\theta$'s and $\kappa$'s
have been here compared as shown in Fig. 4. It is important to emphasize
that the expression used by us as the definition of the heat flux
allows a clear separation between thermodynamic fluxes and forces
and has become standard since the classical work of de Groot and Mazur
\cite{key-19}.

Finally, the coefficients $\zeta_{I}-\zeta_{III}$ in Eq. (7.20) are
related to the so-called Benedicks effect in the presence of a magnetic
field (coefficients of $\vec{E}'$) and the diffusion thermo effect
(coefficients of $\nabla\left(\frac{n_{i}}{n}\right)$ and $\nabla p$),
the standard Dufour coefficients, which we shall label \char`\"{}all
together\char`\"{} and call the $D_{\parallel}$, $D_{\perp}$ and
$D_{s}$. These are absent in Marshall's work.

Indeed, in an entirely analogous fashion the second line of Eq.\,(\ref{37})
can be written as\begin{equation}
\left(\vec{J}_{q}'\right)^{m}=-D_{\parallel}\vec{d}_{ab\parallel}-D_{\perp}\vec{d}_{ab\perp}-D_{s}\vec{d}_{ab\, s}\,,\label{50}\end{equation}
 where using Eqs.\,(\ref{des}) leads to,\begin{equation}
D_{\parallel}=\frac{5}{4}\frac{\left(kT\right)^{2}}{m_{e}}\times0.29n\tau\label{51}\end{equation}
\begin{equation}
D_{\perp}=\frac{5}{4}\frac{\left(kT\right)^{2}}{m_{e}}\frac{n\tau}{\Delta_{1}}\times\left(5.6-66.2x^{2}-0.22x^{4}\right)\label{52}\end{equation}
\begin{equation}
D_{s}=\frac{5}{4}\frac{\left(kT\right)^{2}}{m_{e}}\frac{n\tau}{\Delta_{1}}\times\left(64.7+0.22x^{3}\right)\label{53}\end{equation}
Once more when $\vec{B}=0$, $D_{\perp}=D_{\parallel}$ and we recover
Eq.\,(\ref{39}) identifying $D_{12}$ with $D_{\parallel}$. The
magnitude of these coefficients are shown in Fig. 3.

Let us examine in detail the structure of the diffusive force $\vec{d}_{ij}$
when the calculations are performed in the comoving system so that
$\vec{u}=0$. For the fully ionized gas $\nabla\frac{n_{i}}{n}=0$
and therefore,\begin{equation}
\vec{d}_{ij}=\frac{n_{i}n_{j}}{n}\left(\frac{m_{j}-m_{i}}{\rho}\right)\frac{\nabla p}{p}\simeq\frac{1}{2}\frac{\nabla T}{T}\,,\label{55}\end{equation}
using also the fact that $m_{j}\gg m_{i}$. For this case, as mentioned
above, the heat flux is enhanced by the thermal pressure diffusion
since Eq.\,(\ref{37}) now reads\begin{equation}
\vec{J}_{q}'=-\frac{5}{2}k^{2}T\left\{ \kappa_{\parallel}^{*}\nabla_{\parallel}T+\kappa_{\perp}^{*}\nabla_{\perp}T+\kappa_{s}^{*}\nabla_{s}T\right\} \,,\label{56}\end{equation}
where $\kappa^{*}=\kappa+\frac{D}{2T}$ for each term. Note that,
even in the absence of a magnetic field, the parallel thermal conductivity
is enhanced by a factor $\kappa_{\parallel}^{*}/\kappa_{\parallel}$
of only about $7\%$. The magnitudes of the effective and standard
thermal conductivities are shown in Figs. 4 and 5.

To finish this section we must also emphasize the fact that there
is a magnetic contribution both to mass and energy transport that
arises from the term\begin{equation}
\vec{d}_{ij}=\frac{n_{i}n_{j}}{p\rho}\left(m_{i}e_{j}-m_{j}e_{i}\right)\vec{u}\times\vec{B}\,,\label{57}\end{equation}
which also contains the here neglected effects of an electric field
which, if present, will appear in Eq.\,(\ref{57}) as $\vec{E}'=\vec{E}+\vec{u}\times\vec{B}$.
This modification is rather important when studying the problem of
electrical conduction but we shall leave the details for a future
publication.

\section{concluding remarks}

In this paper we have used the standard Boltzmann equation to study
a fully ionized dilute plasma in a magnetic field. The main results
obtained are explicit formulas for the transport coefficients associated
with mass and heat flux as well as their cross effects. These latter
ones have been completely ignored in the literature; here we stress
their importance for different values of the magnetic field for a
given particle density and temperature under the weak field approximation
so that if $\tau$ is the mean collision time, $\omega_{e}\tau$,
where $\omega_{e}$ is the Larmor frequency for electrons, is not
too large. The full results are shown in the previous section. These
results confirm Balescu's conjecture stating that the values of the
transport coefficients are not very sensitive to the method used in
obtaining them. Also, we obtain an enhanced heat conduction due to
the Dufour cross-effect which leads to an effective thermal conductivity
in all directions {[}Eq.\,(\ref{56})] even in the absence of a magnetic
field.

We could have easily added an external field $\vec{E}$ without any
effort. This would lead to the study of electrical conductivity since
in the expression for $\vec{d}_{ij}$, Eq.\,(\ref{11}), the last
term contains the full Lorentz force $\vec{E}+\vec{u}\times\vec{B}$.
Since the electrical conduction flow is defined as\begin{equation}
\vec{J}_{e}=\sum_{i=a}^{b}\frac{e_{i}}{m_{i}}\vec{J}_{m}\,,\label{58}\end{equation}
 and $\vec{J}_{a}+\vec{J}_{b}=0$,\begin{equation}
\vec{J}_{e}=\frac{m_{a}+m_{b}}{m_{a}m_{b}}e\vec{J_{a}}\,,\label{59}\end{equation}
 thus, using the results for the mass flow, $\vec{J}_{e}$ can be
readily computed from Eq.\,(\ref{59}). We avoided this discussion
because of the intrinsic interest of electrical conduction which deserves
a separate treatment. This has been discussed in another paper \cite{key-45}.

The restriction here to a fully ionized hydrogen plasma is merely
didactic. Taking $Z\neq1$ and setting $n_{a}=\mathcal{X}\left(\vec{r},t\right)\, n$
where $\mathcal{X}\left(\vec{r},t\right)<1$ is a trivial step, all
the collision integrals in the Appendix B can be written in terms
of $\mathcal{X}\left(\vec{r},t\right)$. In such a case $\nabla\frac{n_{a}}{n}\neq0$
and besides the pressure diffusion coefficient there will be a true
Fickian contribution. The reader may pursue this generalization with
no problem whatsoever.

Finally, a word is indispensable concerning the results here obtained
and those obtained by Balescu \cite{key-29} and Braginski \cite{key-7}
using the Fokker-Planck (FP) equation. Balescu uses the Grad-like
moment expansion method to compute the transport coefficients both
for electrons and ions in a plasma for different values of $Z$. As
we mentioned in the previous section the values for the thermal conductivities
of the electrons is rather satisfactory. No comparison is possible
with any cross coefficients since neither he nor Braginski calculate
them. However, a word of caution must be here introduced. Moment methods,
as has been clearly pointed out recently \cite{key-34,key-35}, have
to be dealt with carefully. On the one hand they may be subject to
inconsistencies and secondly, when computing a transport coefficient,
say with N (>5) moments, it is not clear which are the relevant contributions
arising from different orders in the gradients. Although this problem
may arise in the Chapman-Enskog expansion it is automatically taken
care off by Knudsen's parameter. Therefore the different contributions
in successive order in the gradients arise, Navier-Stokes, Burnett
(second order), super-Burnett (third order) and so on. A fair comparison
of the two methods would in principle require the introduction of
such a parameter in the results derived from any moment expansion
to classify the different contributions in terms of the gradient order.
Despite its drawbacks, this method is still used in plasma kinetic
theory calculations (see for example Ref. \cite{key-38})

The results obtained by Braginski are carefully examined by Balescu
so it is not necessary to repeat them here. Let us only add that in
his result the relationship of the FP equation with the tenets of
CIT is not pursued. The fact that the characteristic vector space
to which the solution of the Boltzmann equation pertains is absent
prevents the full discussion of the various cross effects when $\vec{B}\neq0$.
The so-called frictional force introduced in the solution of the FP
equation is artificial, the macroscopic (barycentric) velocity $\vec{u}$
is not a thermodynamic force and according to Curie's principle \cite{key-19}
the only vectorial flows which may couple together are the heat, mass
and charge fluxes. Thus we are convinced that our results are finer
in their content than those arising from the FP equation. We will
not pursue this discussion further since it would lead us away from
the objective here acomplished.

\appendix

\section{integral equations for $\vec{\mathbb{A}}_{j}$ and $\vec{\mathbb{D}}_{j}$}

In this appendix we outline the procedure leading to Eqs.\,(\ref{15})
and (\ref{16}) starting from Eq.\,(\ref{10}). Substitution of Eq.\,(\ref{12})
into Eq.\,(\ref{10}) yields two equations,\begin{eqnarray}
f_{a}^{\left(0\right)}\left(\frac{m_{i}c_{a}^{2}}{2kT}-\frac{5}{2}\right)\vec{c}_{a} & = & -f_{a}^{\left(0\right)}\frac{e_{a}}{m_{a}}\left(\vec{c}_{a}\times\vec{B}\right)\cdot\frac{\partial\vec{\mathbb{A}}_{a}}{\vec{c}_{a}}\nonumber \\
 & - & f_{a}^{\left(0\right)}\frac{m_{a}}{\rho kT}\left[\sum_{j}e_{j}\int d\vec{c}_{j}f_{j}^{\left(0\right)}\vec{\mathbb{A}}_{j}\left(\vec{c}_{j}\times\vec{B}\right)\right]\cdot\vec{c}_{a}\label{a1}\\
 & + & f_{a}^{\left(0\right)}\left[C\left(\vec{\mathbb{A}}_{a}\right)+C\left(\vec{\mathbb{A}}_{a}+\vec{\mathbb{A}}_{b}\right)\right]\,,\nonumber \end{eqnarray}
 an identical equation for species $b$ and two analogous equations
for $\vec{\mathbb{D}}_{a}$ and $\vec{\mathbb{D}}_{b}$ which differ
from Eq.\,(\ref{a1}) in their inhomogeneous term, $\frac{n_{a}}{n}\vec{c}_{a}f_{a}^{\left(0\right)}$.
Thus, the procedure to be applied to Eq.\,(\ref{a1}) is identical
in the three other cases. Substitution of Eq.\,(\ref{13}) into Eq.\,(\ref{a1})
splits each term in this equation into three terms one for each scalar
function $\vec{\mathbb{A}}_{a}^{\left(i\right)}$, $i=1,\,2,\,3$.
Define $M$ as\begin{eqnarray}
M & = & -f_{a}^{\left(0\right)}\frac{e_{a}}{m_{a}}\left(\vec{c}_{a}\times\vec{B}\right)\cdot\frac{\partial}{\partial\vec{c}_{a}}\left[\mathbb{A}_{a}^{\left(1\right)}\vec{c}_{a}+\mathbb{A}_{a}^{\left(2\right)}\left(\vec{c}_{a}\times\vec{B}\right)+\mathbb{A}_{a}^{\left(3\right)}\vec{B}\left(\vec{c}_{a}\cdot\vec{B}\right)\right]\nonumber \\
 &  & -f_{a}^{\left(0\right)}\frac{m_{a}}{\rho kT}\left[\sum_{j}e_{j}\int d\vec{c}_{j}f_{j}^{\left(0\right)}\vec{\mathbb{A}}_{j}\left(\vec{c}_{j}\times\vec{B}\right)\right]\cdot\vec{c}_{a}\,,\label{a2}\end{eqnarray}
 and examine term by term. The first one\begin{equation}
\left(\vec{c}_{a}\times\vec{B}\right)\cdot\frac{\partial}{\partial\vec{c}_{a}}\left(\mathbb{A}_{a}^{\left(1\right)}\vec{c}_{a}\right)=\left(\vec{c}_{a}\times\vec{B}\right)\mathbb{A}_{a}^{\left(1\right)}\,\label{a21}\end{equation}
 since the second term vanishes. For the second one,\begin{eqnarray}
\left(\vec{c}_{a}\times\vec{B}\right)\cdot\frac{\partial}{\partial\vec{c}_{a}}\left[\mathbb{A}_{a}^{\left(2\right)}\left(\vec{c}_{a}\times\vec{B}\right)\right] & = & \mathbb{A}_{a}^{\left(2\right)}\left(\vec{c}_{a}\times\vec{B}\right)\cdot\frac{\partial}{\partial\vec{c}_{a}}\left(\vec{c}_{a}\times\vec{B}\right)\nonumber \\
 &  & +\left(\vec{c}_{a}\times\vec{B}\right)\cdot\left[\left(\vec{c}_{a}\times\vec{B}\right)\frac{\partial\mathbb{A}_{a}^{\left(2\right)}}{\partial\vec{c}_{a}}\right]\,.\label{a22}\end{eqnarray}
 we make use of the vector identity\[
\nabla_{\vec{c}_{a}}\left(\vec{c}_{a}\times\vec{B}\right)=\left(\nabla\vec{c}_{a}\right)\times\vec{B}=\mathbb{I}\times\vec{B}\,,\]
 and thus\[
\left(\vec{c}_{a}\times\vec{B}\right)\cdot\nabla_{\vec{c}_{a}}\left(\vec{c}_{a}\times\vec{B}\right)=\left(\vec{c}_{a}\times\vec{B}\right)\cdot\left(\mathbb{I}\times\vec{B}\right)=\left(\vec{c}_{a}\times\vec{B}\right)\cdot\vec{B}-B^{2}\vec{c}_{a}\]
 where the subscript reminds us that all derivatives are taken with
respect to $\vec{c}_{a}$. The third term in the first line of Eq.\,(\ref{a2})
vanishes.

After exchanging $\left(\vec{c}_{j}\times\vec{B}\right)\cdot\vec{c}_{a}=-\left(\vec{c}_{a}\times\vec{B}\right)\cdot\vec{c}_{j}$
in the second line of Eq.\,(\ref{a2}) we are left with\begin{equation}
f_{a}^{\left(0\right)}\frac{m_{a}}{\rho kT}\left\{ \sum_{j}e_{j}\int d\vec{c}_{j}f_{j}^{\left(0\right)}\vec{c}_{j}\left[\mathbb{A}_{j}^{\left(1\right)}\vec{c}_{j}+\mathbb{A}_{j}^{\left(2\right)}\left(\vec{c}_{j}\times\vec{B}\right)+\mathbb{A}_{j}^{\left(3\right)}\vec{B}\left(\vec{c}_{j}\cdot\vec{B}\right)\right]\right\} \cdot\left(\vec{c}_{a}\times\vec{B}\right)\,,\label{a25}\end{equation}
 and the first term yields\begin{equation}
\sum_{j}e_{j}\int d\vec{c}_{j}f_{j}^{\left(0\right)}\mathbb{A}_{j}^{\left(1\right)}\vec{c}_{j}\vec{c}_{j}=\frac{1}{3}\sum_{j}e_{j}\int d\vec{c}_{j}f_{j}^{\left(0\right)}\mathbb{A}_{j}^{\left(1\right)}c_{j}^{2}\mathbb{I}\,.\label{a26}\end{equation}
 The second one is $\left(\frac{1}{3}\sum_{j}e_{j}\int d\vec{c}_{j}f_{j}^{\left(0\right)}\mathbb{A}_{j}^{\left(1\right)}c_{j}^{2}\mathbb{I}\right)\times\vec{B}$.
However, $\left[\left(\vec{c}_{a}\times\vec{B}\right)\cdot\mathbb{I}\right]\times\vec{B}=\left(\vec{c}_{a}\times\vec{B}\right)\times\vec{B}$
and since the third term vanishes, $\vec{B}\cdot\left(\vec{c}_{a}\times\vec{B}\right)=0$,
we get after the use of the identity\begin{equation}
\int d\vec{c}_{j}\left[c_{j}^{2}-\frac{1}{B^{2}}\left(\vec{c}_{j}\cdot\vec{B}\right)^{2}\right]f_{j}^{\left(0\right)}R\left(c_{j}\right)=\frac{2}{3}\int d\vec{c}_{j}f_{j}^{\left(0\right)}R\left(c_{j}\right)c_{j}^{2}\,,\label{a27}\end{equation}
 for any arbitrary even function of $c_{j}$, $R\left(c_{j}\right)$,
that\begin{eqnarray}
M & = & -f_{a}^{\left(0\right)}\frac{e_{a}}{m_{a}}\left(\vec{c}_{a}\times\vec{B}\right)\mathbb{A}_{a}^{\left(1\right)}+f_{a}^{\left(0\right)}\frac{e_{a}}{m_{a}}\left[B^{2}\vec{c}_{a}-\vec{B}\left(\vec{c}_{a}\cdot\vec{B}\right)\right]\mathbb{A}_{a}^{\left(2\right)}\nonumber \\
 &  & +f_{a}^{\left(0\right)}\frac{m_{a}}{\rho kT}\left(\vec{c}_{a}\times\vec{B}\right)G_{B}^{\left(1\right)}-f_{a}^{\left(0\right)}\frac{m_{a}}{\rho kT}\left[B^{2}\vec{c}_{a}-\vec{B}\left(\vec{c}_{a}\cdot\vec{B}\right)\right]G_{B}^{\left(2\right)}\,,\label{a3}\end{eqnarray}
 where\begin{equation}
G_{B}^{\left(1\right)}=\frac{1}{2}\sum_{j}e_{j}\int d\vec{c}_{j}f_{j}^{\left(0\right)}\mathbb{A}_{j}^{\left(1\right)}\left[c_{j}^{2}-\frac{1}{B^{2}}\left(\vec{c}_{j}\cdot\vec{B}\right)^{2}\right]\,,\label{a4}\end{equation}
 \begin{equation}
G_{B}^{\left(2\right)}=\frac{1}{2}\sum_{j}e_{j}\int d\vec{c}_{j}f_{j}^{\left(0\right)}\mathbb{A}_{j}^{\left(2\right)}\left[c_{j}^{2}-\frac{1}{B^{2}}\left(\vec{c}_{j}\cdot\vec{B}\right)^{2}\right]\,.\label{a5}\end{equation}
 When Eqs.\,(\ref{a3})-(\ref{a5}) are substituted into Eq.\,(\ref{a1})
we get a linear form in terms of the basic vectors $\vec{c}_{a}$,
$\vec{c}_{a}\times\vec{B}$ and $\vec{B}\left(\vec{c}_{a}\cdot\vec{B}\right)$.
Setting the coefficients equal to zero we obtain three equations,
namely,\begin{equation}
\left(\frac{m_{a}c_{a}^{2}}{2kT}-\frac{5}{2}\right)\vec{c}_{a}=\frac{e_{a}B^{2}}{m_{a}}\mathbb{A}_{a}^{\left(2\right)}\vec{c}_{a}-\frac{m_{a}}{\rho kT}B^{2}\vec{c}_{a}G_{B}^{2}+C\left(\mathbb{A}_{a}^{\left(1\right)}\vec{c}_{a}\right)+C\left(\mathbb{A}_{a}^{\left(1\right)}\vec{c}_{a}+\mathbb{A}_{b}^{\left(1\right)}\vec{c}_{b}\right)\,,\label{a6}\end{equation}
 \begin{eqnarray}
0 & = & -\frac{e_{a}}{m_{a}}\mathbb{A}_{a}^{\left(1\right)}\left(\vec{c}_{a}\times\vec{B}\right)+\frac{m_{a}}{\rho kT}\left(\vec{c}_{a}\times\vec{B}\right)G_{B}^{\left(1\right)}+C\left[\mathbb{A}_{a}^{\left(2\right)}\left(\vec{c}_{a}\times\vec{B}\right)\right]\nonumber \\
 &  & +C\left[\mathbb{A}_{a}^{\left(2\right)}\left(\vec{c}_{a}\times\vec{B}\right)+\mathbb{A}_{b}^{\left(2\right)}\left(\vec{c}_{b}\times\vec{B}\right)\right]\,,\label{a7}\end{eqnarray}
 \begin{equation}
0=-\frac{e_{a}}{m_{a}}\mathbb{A}_{a}^{\left(2\right)}\vec{c}_{a}+\frac{m_{a}}{\rho kT}\vec{c}_{a}G_{B}^{\left(2\right)}+C\left(\mathbb{A}_{a}^{\left(3\right)}\vec{c}_{a}\right)+C\left(\mathbb{A}_{a}^{\left(3\right)}\vec{c}_{a}+\mathbb{A}_{b}^{\left(3\right)}\vec{c}_{b}\right)\,.\label{a8}\end{equation}
 Multiplying the third equation by $B^{2}$ and adding it to the first
one yields Eq.\,(\ref{15}) in the text whereas multiplying the second
one by $iB$ and adding it to the first one yields Eq.\,(\ref{16})
in the text. Identical results are obtained for species $b$ and also
for the $\mathbb{D}_{i}^{\left(j\right)}$ functions $i=a,\, b$;
$j=1,\,2,\,3$. The only difference, as we have pointed out above
is in the inhomogeneous term, and instead of the $G$ function we
obtain\begin{equation}
K_{B}^{\left(i\right)}=\frac{1}{2}\sum_{j}e_{j}\int d\vec{c}_{j}f_{j}^{\left(0\right)}\mathbb{D}_{j}^{\left(i\right)}\left[c_{j}^{2}-\frac{1}{B^{2}}\left(\vec{c}_{j}\cdot\vec{B}\right)^{2}\right]\qquad i=1,\,2\,.\label{a9}\end{equation}

\section{collision integrals{*}}

As mentioned in Sect. IV, the evaluation of the transport coefficients
requires the knowledge of twelve coefficients of the series given
in Eq.\,(\ref{19}) and its analog for the $\mathbb{D}_{j}^{\left(k\right)}$
functions, namely\begin{equation}
\alpha_{i}^{\left(m\right)}\equiv a_{i}^{\left(1\right)\left(m\right)}+B^{2}a_{i}^{\left(3\right)\left(m\right)}\qquad i=a,\, b\,,\quad m=0,\,1\,,\label{b1}\end{equation}
 \begin{equation}
\delta_{i}^{\left(m\right)}\equiv d_{i}^{\left(1\right)\left(m\right)}+B^{2}d_{i}^{\left(3\right)\left(m\right)}\qquad i=a,\, b\,,\quad m=0,\,1\,,\label{b2}\end{equation}
 the $m$ superscript denoting the order of the polynomial in Eq.\,(\ref{19})
and\begin{equation}
a_{i}^{\left(m\right)}\equiv a_{i}^{\left(1\right)\left(m\right)}+iBa_{i}^{\left(2\right)\left(m\right)}\,,\label{b3}\end{equation}
 \begin{equation}
d_{i}^{\left(m\right)}\equiv d_{i}^{\left(1\right)\left(m\right)}+iBd_{i}^{\left(2\right)\left(m\right)}\,,\label{b4}\end{equation}
 with the same values for $i$ and $m$. Notice that $\alpha_{i}^{\left(m\right)}=a_{i}^{\left(m\right)}=a_{i}^{\left(1\right)\left(m\right)}$
if $B=0$ and the same relation holds for the $d$'s.

The determination of the twelve independent coefficients in Eqs.\,(\ref{b1})-(\ref{b4})
is achieved using a variational method first introduced by Hirschfelder
\cite{key-21} and afterwards used by several authors \cite{key-12,key-14,key-4,key-5}.
It is based on the linearity property of the linearized Boltzmann
collision kernel. Let $G_{ij}=G_{ij}\left(\vec{w}_{i},\,\vec{w}_{j}\right)$
and $H_{ij}=H_{ij}\left(\vec{w}_{i},\,\vec{w}_{j}\right)$ be any
two functions of the dimensionless velocities $\vec{w}_{i}$ and $\vec{w}_{j}$.
Define\begin{equation}
\left[G_{ij},\, H_{ij}\right]_{ij}\equiv-\frac{1}{n_{i}n_{j}}\int...\int G_{ij}\left(H'_{ij}-H_{ij}\right)f_{i}^{\left(0\right)}f_{j}^{\left(0\right)}g_{ij}\sigma\left(\Omega\right)d\Omega d\vec{w}_{i}d\vec{w}_{j}\,,\label{b5}\end{equation}
 where the subscript $ij$ in the bracket denotes integration over
the variables $\vec{w}_{i}$ and $\vec{w}_{j}$ and the differential
cross section $\sigma\left(\Omega\right)d\Omega=b\, db\, d\epsilon$
for a collision between two particles of species $i$ and $j$ with
an impact parameter $b$($\leq\lambda_{d}$) and orbit inclination
$\epsilon$.

Using the symmetry properties of the kernel in Eq.\,(\ref{b5}),
the fact that $\left[\,\,\,\right]_{ij}$ is a linear operator and
for any two arbitrary functions $K_{i}=K_{i}\left(\vec{w}_{i}\right)$
and $L_{j}=L_{j}\left(\vec{w}_{j}\right)$ the quantity $\left\{ K,\, L\right\} $,\begin{equation}
\left\{ K,\, L\right\} =\sum_{i}n_{i}n_{j}\left[K_{i}+K_{j},\, L_{i}+L_{j}\right]_{ij}\,,\label{b6}\end{equation}
 is such that\begin{equation}
\left\{ K,\, L\right\} =\left\{ L,\, K\right\} \,,\label{b7}\end{equation}
 \begin{equation}
\left\{ K,\, L+M\right\} =\left\{ K,\, L\right\} +\left\{ K,M\right\} \,,\label{b8}\end{equation}
 and\begin{equation}
\left\{ K,\, K\right\} \geq0\,,\label{b9}\end{equation}
 where for obvious reasons the equality follows if and only if $K$
is a linear combination of the collisional invariants. Equation (\ref{b9})
is the basis that sustains the variational procedure. Let $R_{i}$
and $R_{j}$ be the sought solutions to Eq.\,(\ref{15}). Let $t_{i}\left(\vec{w}_{i}\right)$
($i=a,\, b$) be a solution to this equation. Then one can prove that
\cite{key-12,key-14,key-21}\begin{equation}
\sum_{i=a}^{b}\int d\vec{w}_{i}I_{i}\left(\vec{w}_{i}\right):\, t_{i}\left(\vec{w}_{i}\right)=\left\{ t_{i},\, t_{i}\right\} \leq\left\{ R_{i},\, R_{i}\right\} \,,\label{b10}\end{equation}
 where $I_{i}\left(\vec{w}_{i}\right)$, the inhomogeneous term, is
known. Thus the trial function $t$ must be chosen so to maximize
the collision integral $\left\{ R_{i},\, R_{i}\right\} $. For the
trial function we propose that\begin{equation}
\vec{t}_{i}=\vec{c}_{i}\sum_{p=0}^{M}a_{p}^{\left(p\right)}S_{3/2}^{\left(p\right)}\left(c_{i}\right)\,,\label{b11}\end{equation}
 so that $M$ characterizes the order of the approximation. The left
side of Eq.\,(\ref{b10}) can be readily integrated after substitution
of Eq.\,(\ref{b11}) and use of Eq.\,(\ref{20}). This leads to
the result that\begin{equation}
15kT\left(\frac{n_{a}}{m_{a}}\alpha_{a}^{\left(1\right)}+\frac{n_{b}}{m_{b}}\alpha_{b}^{\left(1\right)}\right)=\left\{ t_{0},\, t_{i}\right\} \,.\label{b12}\end{equation}
 We now choose that, to a first approximation, $m=0,\,1$ so that\begin{equation}
K_{i}=L_{i}=\alpha_{i}^{\left(0\right)}\vec{c}_{i}+\alpha_{i}^{\left(1\right)}S_{3/2}^{\left(1\right)}\vec{c}_{i}\qquad i=a,\, b\,.\label{b13}\end{equation}
 It is readily shown that for a binary mixture Eq.\,(\ref{b6}) gives
\begin{eqnarray}
\left\{ K,\, L\right\}  & = & 2n_{a}^{2}\left[K_{a},\, L_{a}\right]_{aa}+2n_{b}^{2}\left[K_{b},\, L_{b}\right]_{bb}+2n_{a}n_{b}\left\{ \left[K_{a},\, L_{a}\right]_{ab}\right.\nonumber \\
 & + & \left.\left[K_{a},\, L_{b}\right]_{ab}+\left[K_{b},\, L_{a}\right]_{ab}+\left[K_{b},\, L_{b}\right]_{ab}\right\} \,.\label{b14}\end{eqnarray}
 Equations (\ref{b13}) and (\ref{b14}), after a lengthy but straightforward
algebra, lead to a system of algebraic equations for $\alpha_{a}^{\left(0\right)}$,
$\alpha_{b}^{\left(0\right)}$, $\alpha_{a}^{\left(1\right)}$ and
$\alpha_{b}^{\left(1\right)}$ whose coefficients are fourteen collision
integrals which must be explicitly evaluated. This is an exercise,
tedious but rather straightforward, which yields the following results
when $m_{b}\gg m_{a}$ \cite{key-14},\[
\left[\vec{w}_{a},\,\vec{w}_{a}\right]_{aa}=\left[\vec{w}_{b},\,\vec{w}_{b}\right]_{bb}=0\,,\]
 \[
\left[\vec{w}_{a},\,\vec{w}_{a}\right]_{ab}=\varphi\,,\]
 \[
\left[\vec{w}_{b},\,\vec{w}_{b}\right]_{ab}=M_{1}\varphi\,,\]
 \[
\left[\vec{w}_{a},\,\vec{w}_{a}\left(w_{a}^{2}-\frac{5}{2}\right)\right]_{ab}=\frac{3}{2}\varphi\,,\]
 \[
\left[\vec{w}_{b},\,\vec{w}_{b}\left(w_{b}^{2}-\frac{5}{2}\right)\right]_{ab}=\frac{3}{2}M_{1}^{2}\varphi\,,\]
 \begin{equation}
\left[\vec{w}_{a},\,\vec{w}_{b}\right]_{ab}=-\sqrt{M_{1}}\varphi\,,\label{ces}\end{equation}
 \[
\left[\vec{w}_{i},\,\vec{w}_{i}\left(w_{i}^{2}-\frac{5}{2}\right)\right]_{ii}=0\quad i=a,\, b\,,\]
 \[
\left[\vec{w}_{a}\left(w_{a}^{2}-\frac{5}{2}\right),\,\vec{w}_{a}\left(w_{a}^{2}-\frac{5}{2}\right)\right]_{ab}=\frac{13}{4}\varphi\,,\]
 \[
\left[\vec{w}_{b}\left(w_{b}^{2}-\frac{5}{2}\right),\,\vec{w}_{b}\left(w_{b}^{2}-\frac{5}{2}\right)\right]_{ab}=\frac{15}{2}M_{1}\varphi\,,\]
 \[
\left[\vec{w}_{a}\left(w_{a}^{2}-\frac{5}{2}\right),\,\vec{w}_{b}\left(w_{b}^{2}-\frac{5}{2}\right)\right]_{ab}=-\frac{27}{4}M_{1}^{3/2}\varphi\,,\]
 \[
\left[\vec{w}_{a},\,\vec{w}_{b}\left(w_{b}^{2}-\frac{5}{2}\right)\right]_{ab}=-\frac{3}{2}M_{1}^{3/2}\varphi\,,\]
 \[
\left[\vec{w}_{a}\left(w_{a}^{2}-\frac{5}{2}\right),\,\vec{w}_{b}\right]_{ab}=-\frac{3}{2}M_{1}^{1/2}\varphi\,,\]
\[
\left[\vec{w}_{a}\left(w_{a}^{2}-\frac{5}{2}\right),\,\vec{w}_{a}\left(w_{a}^{2}-\frac{5}{2}\right)\right]_{aa}=\sqrt{2}\varphi\,,\]
\[
\left[\vec{w}_{b}\left(w_{b}^{2}-\frac{5}{2}\right),\,\vec{w}_{b}\left(w_{b}^{2}-\frac{5}{2}\right)\right]_{bb}=\sqrt{2M_{1}}\varphi\,,\]
 where \[
M_{1}=\frac{m_{a}}{m_{a}+m_{b}}\,,\]
 and\begin{equation}
\varphi\equiv\frac{1}{n\tau}=\frac{e^{4}\psi}{4\left(2\pi\right)^{3/2}\sqrt{m_{a}}\epsilon_{0}^{2}\left(kT\right)^{3/2}}\,,\label{b15}\end{equation}
 where all symbols are defined in Eq.\,(\ref{42}) and $\psi$ in
Eq.\,(\ref{43}). The mean free time $\tau$ is defined here as the
inverse of the collision integral $\varphi$; we stress here that
the logarithmic function is a consequence of the evaluation of the
collision integrals. Using these collision integrals and Eq.\,(\ref{b15})
the resulting algebraic system, too long to be given here, is maximized
to satisfy Eq.\,(\ref{b10}) making use of the subsidiary conditions
Eqs.\,(\ref{25}) and (\ref{26}). After the system is solved one
finds that\[
\alpha_{a}^{\left(0\right)}=2.94\tau=a_{a}^{\left(1\right)\left(0\right)}+B^{2}a_{a}^{\left(3\right)\left(0\right)}\,,\]
 \begin{equation}
\alpha_{a}^{\left(1\right)}=1.96\tau=a_{a}^{\left(1\right)\left(1\right)}+B^{2}a_{a}^{\left(3\right)\left(1\right)}\,,\label{alfas}\end{equation}
 \[
\alpha_{b}^{\left(1\right)}=107\tau=a_{b}^{\left(1\right)\left(1\right)}+B^{2}a_{b}^{\left(3\right)\left(1\right)}\,.\]
 Here we have introduced the restrictive (but completely unnecessary)
condition that the plasma is fully ionized, $n_{a}=n_{b}=\frac{n}{2}$
so by Eq.\,(\ref{25}) $\alpha_{a}^{\left(0\right)}=\alpha_{b}^{\left(0\right)}$.
Also $m_{b}\gg m_{a}$ ($m_{p}=1836\, m_{e}$) so that $M_{1}\sim\frac{m_{e}}{m_{p}}$.

An identical calculation for the $\mathbb{D}_{j}^{\left(k\right)}$
functions leads to the results that\[
\delta_{a}^{\left(0\right)}=-\delta_{b}^{\left(0\right)}=1.2\tau=d_{a}^{\left(1\right)\left(0\right)}+B^{2}d_{a}^{\left(3\right)\left(0\right)}\,,\]
 \begin{equation}
\delta_{a}^{\left(1\right)}=0.29\tau=d_{a}^{\left(1\right)\left(1\right)}+B^{2}d_{a}^{\left(3\right)\left(1\right)}\,,\label{deltas}\end{equation}
 \[
\delta_{b}^{\left(1\right)}=1.5\times10^{-3}\tau=d_{b}^{\left(1\right)\left(1\right)}+B^{2}d_{b}^{\left(3\right)\left(1\right)}\,.\]
 The solution to Eq.\,(\ref{16}) is slightly more sophisticated
due to the presence of the two field dependent terms in its right
side. The method followed is apparently due to N. Davison \cite{key-12,key-25}.
One proposes a trial function $\zeta_{i}$ and constructs a functional
$D\left(\zeta_{i}\right)$ by multiplying Eq.\,(\ref{16}) by $\zeta_{i}$
and integrating over $d\vec{c}_{i}$ and summing for $i=a,\, b$.
It then follows that the term containing $G$ vanishes on account
of the subsidiary conditions, Eqs.\,(\ref{c1}). One then seeks a
solution satisfying the extremal condition\[
\delta D\left(\zeta_{i}\right)=0\,,\]
 consistent with Eqs.\,(\ref{c1}), after $\varphi_{i}^{\left(1\right)}$
is substituted by Eq.\,(\ref{12}). Using then the invariance of
the collision kernels under the exchange of velocities $\vec{c}_{j}\rightarrow\vec{c}_{j}'$,
$\vec{c}_{i}\rightarrow\vec{c}_{i}'$ and microscopic reversibility,
one gets an equation for $\zeta_{i}$ consistent with Eqs.\,(\ref{c1})
which is of the form\[
f_{i}^{\left(0\right)}\left(\frac{m_{i}c_{i}^{2}}{2kT}-\frac{5}{2}\right)\vec{c}_{i}=2iB\frac{e_{i}}{m_{i}}f_{i}^{\left(0\right)}\vec{c}_{i}\zeta_{i}+f_{i}^{\left(0\right)}\left[C\left(\vec{c}_{i}\zeta_{i}\right)+C\left(\vec{c}_{i}\zeta_{i}+\vec{c}_{j}\zeta_{j}\right)\right]\,.\]
 If we now choose for $\zeta_{i}$\[
\zeta_{i}=\sum_{m=0}^{M}a_{i}^{\left(m\right)}S_{3/2}^{\left(m\right)}\left(c_{i}\right)\,,\]
 and substitute back into the equation for $D\left(\zeta_{i}\right)$,
the left side is readily integrated leading to the result that\[
5kT\sum_{i}\frac{n_{i}}{m_{i}}a_{i}^{\left(1\right)}=-3iBkT\sum_{i}\frac{e_{i}n_{i}}{m_{i}^{2}}\left[2\left(a_{i}^{\left(0\right)}\right)^{2}+5\left(a_{i}^{\left(1\right)}\right)^{2}\right]-\frac{1}{4}\left\{ K,\, L\right\} \,,\]
 where $\left\{ K,\, L\right\} $ is given by Eq.\,(\ref{b14}).
The remaining of the procedure is the same as in the previous case
so that after maximizing the resulting set of algebraic equations,
solving them, and separating real and imaginary parts {[}see Eqs.\,(\ref{b3})-(\ref{b4})]
one gets that

\[
\text{Re}\left[a_{a}^{(0)}\right]=\frac{\tau}{\Delta_{1}}\left(56.3-662x^{2}-2.25x^{4}\right)\]
\[
\text{Im}\left[a_{a}^{(0)}\right]=\frac{\tau}{\Delta_{1}}\left(647x+2.2x^{3}\right)\]
\begin{equation}
\text{Re}\left[a_{a}^{(1)}\right]=\frac{\tau}{\Delta_{1}}\left(37.5+2147x^{2}+7.3x^{4}\right)\label{aes}\end{equation}
\[
\text{Im}\left[a_{a}^{(1)}\right]=\frac{\tau}{\Delta_{1}}\left(206x+2649x^{3}+9x^{5}\right)\]
\[
\text{Re}\left[a_{b}^{(1)}\right]=\frac{\tau}{M_{1}\Delta_{1}}\left(1.12+121.2x^{2}+154.4x^{4}\right)\]
\[
\text{Im}\left[a_{b}^{(1)}\right]=-\frac{\tau}{M_{1}\Delta_{1}}\left(0.06x+7x^{3}+9x^{5}\right)\]
 where $\Delta_{1}$ is defined in Eq.\,(\ref{48}) and $x=\frac{eB}{m_{e}}=1.76\times10^{11}\tau B$
if $B$ is given in teslas.

By a completely identical procedure,\[
\text{Re}\left[d_{a}^{(0)}\right]=\frac{\tau}{\Delta_{1}}\left(22.8+55.3x^{2}+0.19x^{4}\right)\]
\[
\text{Im}\left[d_{a}^{(0)}\right]=\frac{\tau}{\Delta_{1}}\left(234x+331.9x^{3}+1.12x^{5}\right)\]
\begin{equation}
\text{Re}\left[d_{a}^{(1)}\right]=\frac{\tau}{\Delta_{1}}\left(5.6-66.2x^{2}-0.22x^{4}\right)\label{des}\end{equation}
\[
\text{Im}\left[d_{a}^{(1)}\right]=\frac{\tau}{\Delta_{1}}\left(64.7x+0.22x^{3}\right)\]
\[
\text{Re}\left[d_{b}^{(1)}\right]=\frac{\tau}{\Delta_{1}}\left(0.03-4.1x^{2}-0.22x^{4}\right)\]
\[
\text{Im}\left[d_{b}^{(1)}\right]=\frac{\tau}{\Delta_{1}}\left(.66x-3.6x^{3}\right)\]

Notice that consistently in both cases, if $B=0$ the real parts of
these equations agree with the results given in Eqs.\,(\ref{aes})
and (\ref{des}). These equations are those needed to evaluate all
the transport coefficients appearing in the text.

{*} The authors are deeply grateful to Prof. Leonardo Dagdug for checking
all the values of the collision integrals listed in this appendix
and for his aid in all the rather involved algebraic steps involved
in these calculations.

\section{subsidiary conditions}

In Sect. II it is asserted that Eq.\,(\ref{max}) is a solution to
the homogeneous part of Eq.\,(\ref{7}). This statement has a deeper
significance. In fact, as it is well known any linear combination
of the conserved variables $m_{i}$, $m_{i}\vec{v}_{i}$ and $\frac{1}{2}m_{i}v_{i}^{2}$
($i=a,\, b$) is a solution to homogeneous part. This leads to a local
distribution which has five undetermined parameters, all depending
of position $\vec{r}$ and time $t$. To uniquely determine them one
assumes that the five local variables $n\left(\vec{r},t\right)$,
$\vec{u}\left(\vec{r},t\right)$ and $\varepsilon\left(\vec{r},t\right)$
are determined by $f_{i}^{\left(0\right)}$ and \emph{not} by the
full distribution function $f_{i}\left(\vec{r},\vec{v}_{i},t\right)$.
This is the way the local equilibrium assumption is forced into the
kinetic formalism thus implying that, for a multicomponent system,\begin{equation}
\sum_{i}\int f_{i}^{\left(0\right)}\varphi_{i}^{\left(n\right)}\left\{ \begin{array}{c}
m_{i}\\
m_{i}\vec{c}_{i}\\
\frac{1}{2}m_{i}c_{i}^{2}\end{array}\right\} d\vec{c}_{i}=0\qquad n\geq1\,,\label{c1}\end{equation}
 where $\varphi_{i}^{\left(n\right)}$ is the $n^{th}$ coefficient
in the Knudsen expansion of $f_{i}$ {[}Eq.\,(\ref{9})]. Equations
(\ref{c1}) are the so-called subsidiary conditions that all $\varphi_{i}$'s
must obey.

\newpage{} \underbar{\large Figure Captions}{\large \par}

\begin{lyxlist}{00.00.0000}
\item [{Fig.\,1.\:}] The Soret coefficients ($S_{\parallel}$, solid;
$S_{\perp}$, dotted; $S_{s}$, dashed) for $n=10^{21}m^{-3}$ and
$T=10^{7}K$. 
\item [{Fig.\,2.\:}] The behavior of the three thermal conductivities
($\kappa_{\parallel}$, solid; $\kappa_{\perp}$, dotted; $\kappa_{s}$,
dashed) for $n=10^{21}m^{-3}$ and $T=10^{7}K$. 
\item [{Fig.\,3.\:}] The behavior of the thermal diffusion coefficients
$D$ ($D_{\parallel}$, solid; $D_{\perp}$, dotted; $D_{s}$, dashed)
consistent with the condition $x<1$ for $n=10^{21}m^{-3}$ and $T=10^{7}K$. 
\item [{Fig.\,4.\:}] The effective thermal conductivity $\kappa_{\perp}^{*}$
compared with $\kappa_{\perp}$ for $n=10^{21}m^{-3}$ and $T=10^{7}K$. 
\item [{Fig.\,5.\:}] The effective thermal conductivity $\kappa_{s}^{*}$
compared with $\kappa_{s}$ for $n=10^{21}m^{-3}$ and $T=10^{7}K$. 
\end{lyxlist}

\newpage{} %
\begin{figure}
\includegraphics[width=7in,height=7in]{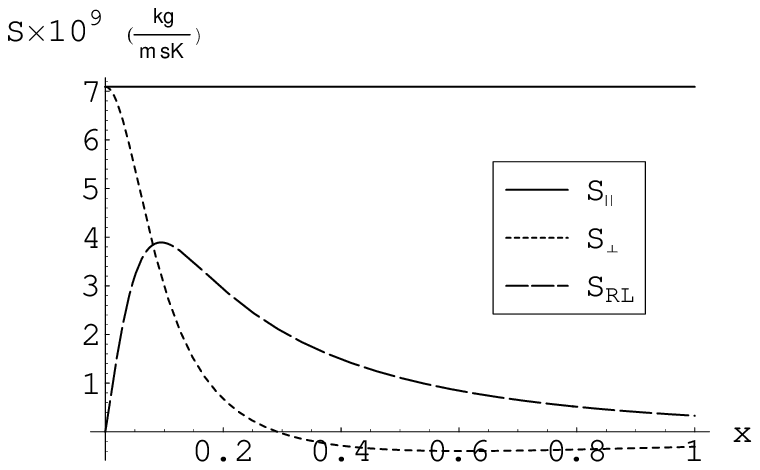}

Fig. 1

\label{fig:F1} 
\end{figure}

\newpage{} %
\begin{figure}
\includegraphics[width=7in,height=7in]{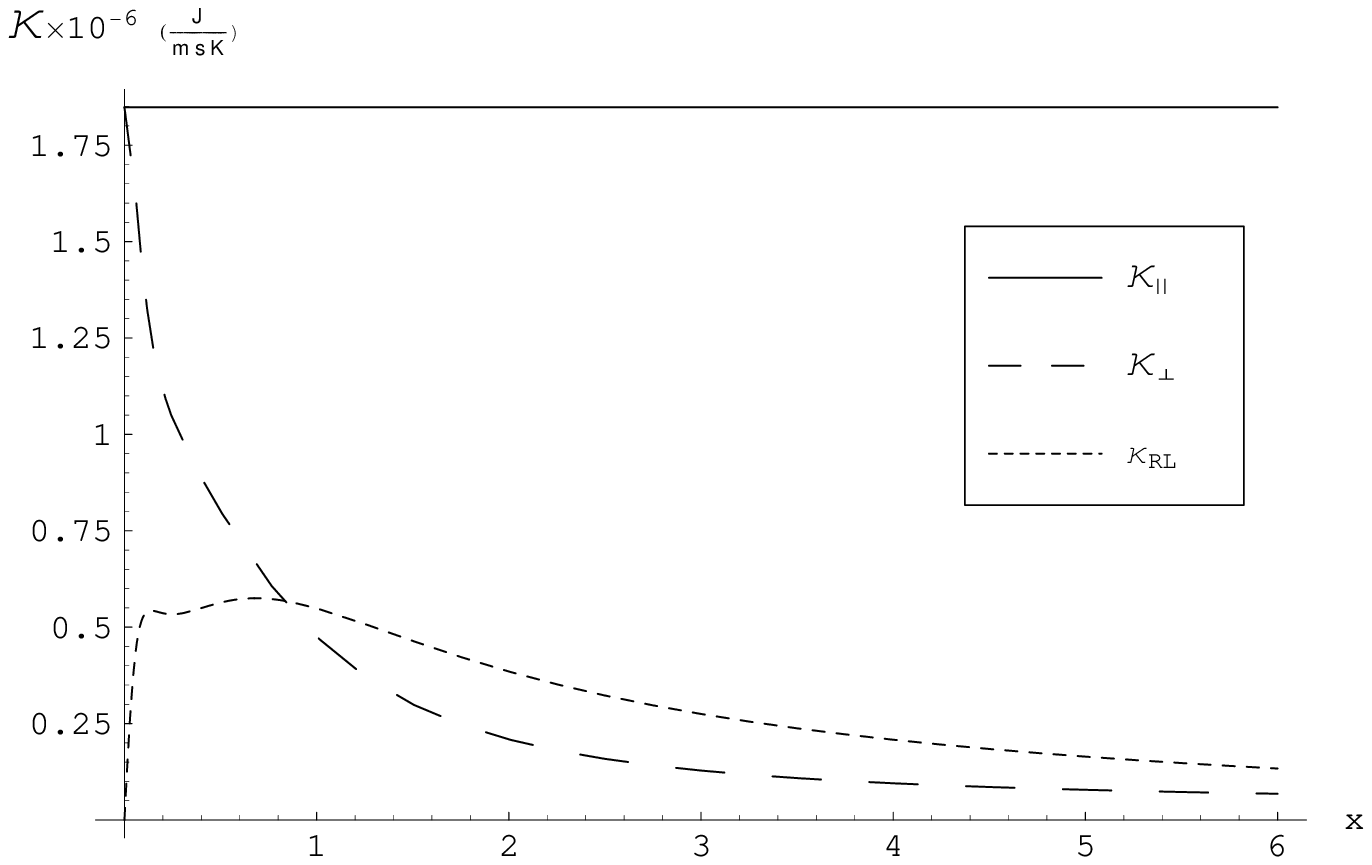}

Fig. 2

\label{fig:F2} 
\end{figure}

\newpage{} %
\begin{figure}
\includegraphics[width=7in,height=7in]{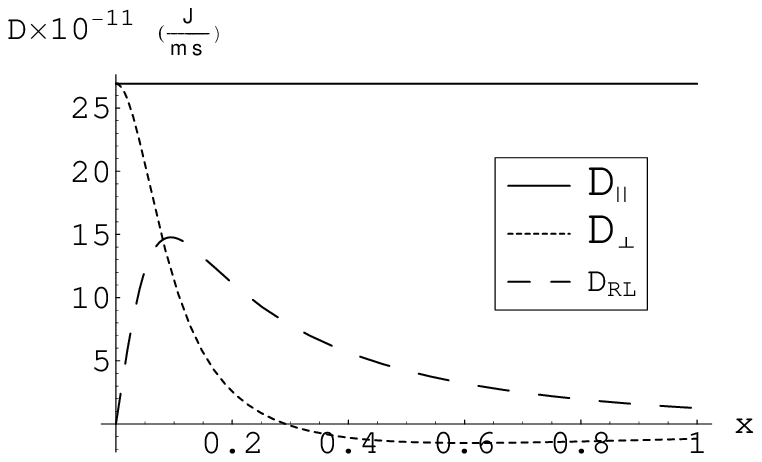}

Fig. 3

\label{fig:F3} 
\end{figure}

\newpage{} %
\begin{figure}
\includegraphics[width=7in,height=7in]{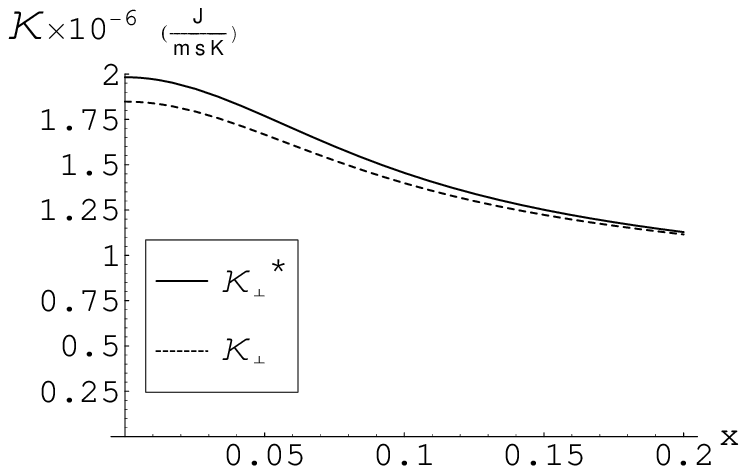}

Fig. 4

\label{fig:F4} 
\end{figure}

\newpage{} %
\begin{figure}
\includegraphics[width=7in,height=7in]{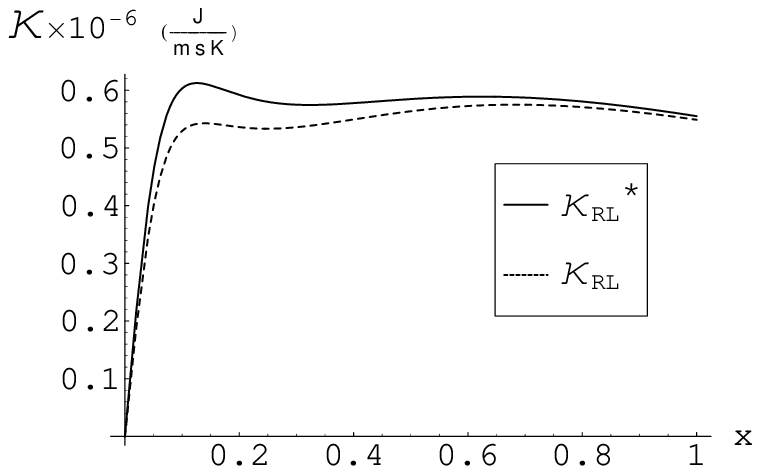}

Fig. 5

\label{fig:F5} 
\end{figure}

\end{document}